\documentclass{elsart}

\usepackage{graphicx}

\usepackage{amssymb}

\begin{document}

\begin{frontmatter}



\title{Reply to ``Remarks on the simulation of Cl electrosorption
on Ag(100) reported in Electrochimica Acta 50 (2005) 5518"}


\author[a,b,c]{P.~A.\ Rikvold\corauthref{cor1}}
\ead{rikvold@scs.fsu.edu}
\author[d]{Th.~Wandlowski}
\author[a,b]{I.\ Abou Hamad\thanksref{label1}}
\author[e]{S.~J.\ Mitchell}
\author[a,b,f]{G.~Brown}
\address[a]{Center for Materials Research and Technology and 
Department of Physics, Florida State University, Tallahassee, FL
32306-4350, USA}
\address[b]{School of Computational Science, Florida State
University, Tallahassee, FL 32306-4120, USA}
\address[c]{National High Magnetic Field Laboratory, Tallahassee,
FL 32310, USA}
\address[d]{Department of Bio- and Nanosystems IBN 3, and
Centre of Nanoelectronic Systems, CNI, Research Centre J{\"u}lich,
52425 J{\"u}lich, Germany}
\address[e]{Center for Simulational Physics and Department of
Physics and Astronomy, The University of Georgia, Athens, GA
30602-2451, USA}
\address[f]{Center for Computational Sciences, Oak Ridge National
Laboratory, Oak Ridge, TN 37831, USA}
\corauth[cor1]{Corresponding author. 
Tel.: +1 850 644 6011; fax: 1 850 644 0098}
\thanks[label1]{Current address: Department of Physics and Astronomy,
Mississippi State University, Mississippi State, MS 39762, USA}

\begin{abstract}
We reply to the remarks by L{\'a}ng and Hor{\'a}nyi [Electrochim.\
Acta Vol. (2006) page] on the meaning of the notion of
``electrosorption valency" used in I.\ Abou Hamad {\it et al.\/}, 
Electrochim.\ Acta 50 (2005) 5518. 
It is concluded that, contrary to the assertion of L{\'a}ng and
Hor{\'a}nyi, the magnitude of the 
current in the external circuit upon adsorption of
an ion of charge $ze$ with partial charge transfer is indeed given
by an electrosorption valency $\gamma$ such that
$| \gamma e| < |ze|$. We believe the conclusion of 
L{\'a}ng and Hor{\'a}nyi to the contrary is the result of an
excessively severe charge-neutrality requirement. 
\end{abstract}

\end{frontmatter}

\section{Introduction}
\label{sec:int}

In their interesting discussion article \cite{LANG06B}, L{\'a}ng and
Hor{\'a}nyi (LH) question the concept of {\it
electrosorption valency\/} \cite{VETT72A,VETT72B} as interpreted 
in our recent paper on Cl electrosorption on Ag(100) \cite{HAMA05B}. 
As we understand their argument, its
central point is the requirement of charge neutrality in the
solution phase and {\it separately\/} in a surface layer consisting
of the specifically adsorbed ions (traditionally known as the 
Inner Helmholtz Layer, IHL) and the 
adjoining part of the metal in the working electrode. These two
separate conditions, when applied simultaneously, lead them to the
conclusion that the charge transported through the external
circuit due to the specific adsorption of an ion of charge $ze$
must be $ze$, regardless of possible partial discharge of the
adsorbate, unless coadsorption of ions of opposite charge takes place. 

We argue that this double application of the charge-neutrality
requirement overconstrains the problem and leads to an erroneous
conclusion. The correct region over which charge neutrality should
be applied must include the parts of the solution phase close to
the IHL, traditionally known as the Outer Helmholtz layer (OHL) and
the diffuse double layer (DDL). See Fig.~\ref{fig:one} for 
definitions of these terms. 
This polarized region, representing the
half-cell of the working electrode, is separated from the rest of
the system by fictitious bounding surfaces inside the uniform,
macroscopically uncharged bulk media (electrolyte and metal,
respectively). This schematic partitioning of the system
removes the need to consider the counter electrode 
explicitly in calculating the charge transport. 
A completely analogous condition of charge neutrality is 
independently obeyed at the counter electrode.
The current through the working electrode into the external circuit
induces a current of opposite sign, but with the same magnitude, through the
counter electrode. For instance, if chloride adsorbs at the working
electrode, a (partial) negative charge will be released into the
external circuit, while a corresponding negative charge will, on
average, be injected
through the counter electrode. In the present case, this latter charge
will most probably transform a corresponding amount of
H$^+$ into $\half$H$_2$, thus maintaining  
electroneutrality in the entire system. 

Our conclusion, for which we argue below, is that the expressions
for electrosorption valency and adsorbate dipole moment used in
Ref.~\cite{HAMA05B} correspond to the approximations of excess
supporting electrolyte, as well as to 
identifying the electrosorption valency with the negative of the
partial charge-transfer coefficient. While not exact, these are 
reasonable approximations, supported {\it a posteriori\/}
by the good agreement between the numerical results for our model system
and the experimental adsorption isotherms.  
In the process we reconfirm Vetter and Schultze's relation between
the electrosorption valency and the current in the external circuit
\cite{VETT72B}. 

Details of our arguments, including the approximations used in
Ref.~\cite{HAMA05B} (and also in Ref.~\cite{HAMA03}), are given 
below. In Sec.~\ref{sec:ESV} we discuss the definition 
of the electrosorption valency; in Sec.~\ref{sec:Q} we discuss its 
relation to the current in the external circuit; and in Sec.~\ref{sec:P} 
we obtain its relation to the surface dipole moment. 
Our conclusions are summarized in Sec.~\ref{sec:C}. 

\section{Electrosorption Valency}
\label{sec:ESV}

The electrosorption valency 
was thermodynamically formalized by Vetter and Schultze 
\cite{VETT72A,VETT72B} to account for the current in the
external circuit during electrosorption with partial
charge transfer \cite{LORE61}. 
Abbreviated derivations have been presented later
(see, e.g., Ch.\ 18 of Ref.~\cite{SCHM96}), but we believe the most
detailed treatment is given in Refs.~\cite{VETT72A,VETT72B}, as
supplemented by Refs.~\cite{SCHU73,VETT74,SCHU76B}. 
Results of these early papers were recently summarized by
Schultze and Rolle \cite{SCHU97,SCHU03}. 

\begin{figure}[tbp]
\begin{center}
\includegraphics[angle=0,width=0.60\textwidth]{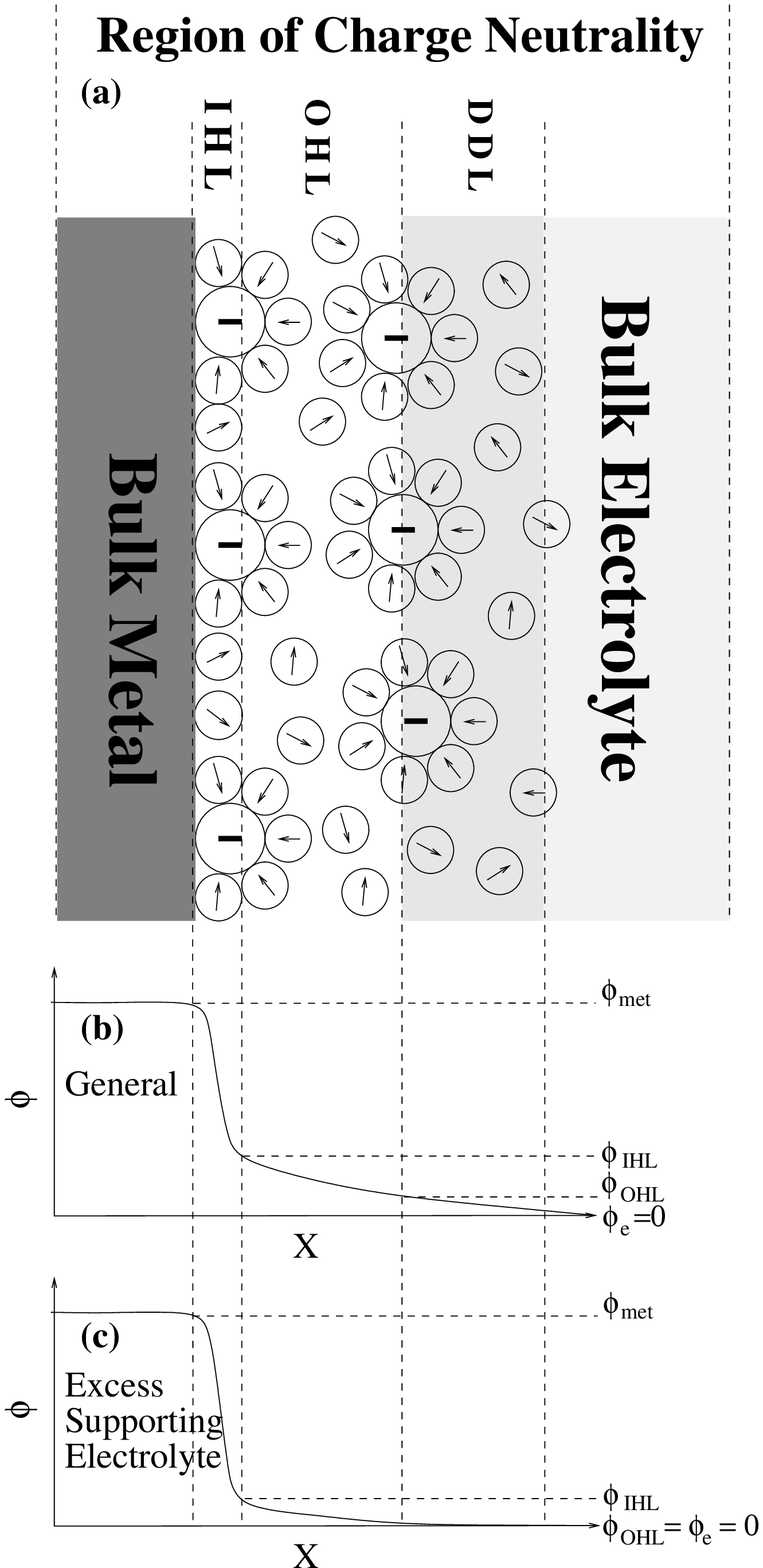}
\end{center}
\caption[]{
Schematic picture of the interface region at the working electrode
{\bf (a)}, the electrostatic 
potential $\phi$ as a function of distance $x$ perpendicular to the 
surface in the general case {\bf (b)} and in the case of excess
supporting electrolyte {\bf (c)}. Here, IHL stands for the inner
Helmholtz layer, OHL for the outer Helmholtz layer, and DDL for the
diffuse double layer. Large circles with a minus sign 
represent the adsorbate ions, and small circles with an arrow
indicating a dipole moment represent water molecules. 
After Ref.~\protect\cite{VETT72A}. 
}
\label{fig:one}
\end{figure}

Vetter and Schultze consider an electrosorption reaction for an ion
of valence $z$ with the possible charge transfer of a noninteger 
number $\lambda$ of electrons, described by the equation
\cite{VETT72A,VETT72B,SCHU73,VETT74,SCHU76B} 
\begin{equation}
\nu \mathrm{M}\!-\!\mathrm{OH}_2 + \mathrm{S}^z \! \cdot \! \mathrm{aq} 
\rightleftharpoons
\mathrm{M}\!-\!\mathrm{S}^{z+\lambda} + \lambda e^-(\mathrm{met}) + 
\nu \mathrm{H}_2\mathrm{O}\! \cdot \! \mathrm{aq} 
\;.
\label{eq:ads}
\end{equation}
To avoid extrathermodynamic complications arising from considering
the microscopic structure of the DDL \cite{FRUM74}, 
we here restrict ourselves to the case of
excess supporting electrolyte, so that the potential at the OHL, 
$\phi_\mathrm{OHL}$, is the same as that in the bulk electrolyte, 
$\phi_\mathrm{e}$, 
\cite{VETT72A,VETT72B,VETT74}, which we can define equal to zero
without loss of generality. See Fig.~\ref{fig:one}. 
This is indeed the condition corresponding to the experiments reported
in Refs.~\cite{HAMA05B,HAMA03}. The electrode potential is then  
$E = \phi_\mathrm{met} + \mathrm{cst.}$, where $\phi_\mathrm{met}$
is the potential of the metal. Thus, in the case of excess
supporting electrolyte, differentiation
with respect to $E$ is the same as with respect to 
$(\phi_\mathrm{met} - \phi_\mathrm{OHL})$. 
The excess of supporting electrolyte also reduces the concentration
of adsorbate ions S$^z$ in the DDL, while a relatively low concentration
of adsorbate ions relative to the solvent reduces their concentration
in the OHL relative to the IHL. In this approximation we can
therefore replace the total surface excess of adsorbate ions,
$\Gamma_\mathrm{S}$, by the surface coverage $\theta$ of the specifically
adsorbed, partially discharged
species S$^{z+\lambda}$ in the IHL \cite{VETT72A}. 
The coverage is defined as the number of specifically adsorbed ions per
adsorption site on the surface. In what follows, we shall use $\theta$
as an approximation for the more general $\Gamma_\mathrm{S}$. 

In Ref.~\cite{VETT72A}, the electrosorption valency is defined as 
\begin{equation}
\gamma = 
\left( \frac{\partial \mu_\mathrm{S} }{\partial E} \right)_{\theta}
\;,
\label{eq:gam}
\end{equation}
where $\mu_\mathrm{S}$ 
is the {\it chemical\/} potential of S$^z$ in the bulk
solution. From this and a general adsorption isotherm corresponding
to the reaction equation (\ref{eq:ads}), they obtain the relation
\begin{equation}
\gamma = \gamma_\mathrm{PZC} 
-
\frac{1}{e} \int_{E_\mathrm{PZC}}^E 
\left( \frac{\partial C_\mathrm{D}}{\partial \theta} \right)_E
\mathrm{d} E'
\;,
\label{eq:gam2}
\end{equation}
where $e$ is the elementary charge unit and $C_\mathrm{D}$ is the
capacitance of the compact double layer (metal vs.\ IHL). The
subscript PZC refers to the Potential of Zero Charge. The value of 
$\gamma$ at the PZC is 
\begin{equation}
\gamma_\mathrm{PZC} 
= 
gz - \lambda(1-g) + \kappa_\mathrm{ad} - \nu \kappa_\mathrm{w} 
\;.
\label{eq:gam3}
\end{equation}
Here, $\kappa_\mathrm{ad}$ and $\kappa_\mathrm{w}$ refer to the
effects of the dipole moments of the adsorbate and water,
respectively, and can usually be neglected for inorganic, aqueous
electrolytes. The factor 
$g =
(\phi_\mathrm{IHL}-\phi_\mathrm{OHL})/
(\phi_\mathrm{met}-\phi_\mathrm{OHL})$ is the ratio of the
potential difference between the adsorbate and the OHL to that
between the metal and the OHL and is frequently of the order of 0.2. 
A simple interpretation of
the schematic Fig.~\ref{fig:one} with a near-linear potential
profile would lead to the traditional
interpretation of $g$ as a purely geometric
factor. For more realistic microscopic models that lead to a nonlinear
potential profile, such as illustrated in Fig.~\ref{fig:one},
the connection of $g$ to the geometry of the interface region is less clear. 

Equations (\ref{eq:gam2}) and~(\ref{eq:gam3}) reveal two important
approximations in our Refs.~\cite{HAMA05B,HAMA03}. The first is
that the integral in Eq.~(\ref{eq:gam2}) is approximated by a
linear function in the coverage $\theta$. This is reasonable since
any changes in $\gamma$ with $E$ are likely to be largely due to
the increased crowding on the surface. 

The second approximation, which is expressed in Sec.~2.2 of our 
Ref.~\cite{HAMA05B} (and also quoted verbatim in the fifth paragraph
of LH \cite{LANG06B}), is that we take $g=0$. This leads to
Lorentz' approximation $\gamma_\mathrm{PZC} = - \lambda$
\cite{LORE61} and corresponds to the situation that the full
potential drop happens between the bulk metal and the IHL. A somewhat 
better approximation is probably the minimum value for water, 
$g_\mathrm{min} \approx 0.16$, obtained by Schultze and Koppitz
\cite{SCHU76B}. 

\section{Surface Charge and Potentiostatic Current}
\label{sec:Q}

The crucial problem of the interpretation of $\gamma$ in terms of
the current in the external circuit is treated by Vetter and  Schultze
in Ref.~\cite{VETT72B}. The central point is that even a
partial discharge corresponding to $\lambda \neq 0$ merely
corresponds to a {\it redistribution of charge in the interface
region\/}. Therefore, all currents in the circuit are {\it
capacitive\/}, corresponding to changes in the excess charge
density on the metal, $q_\mathrm{met}$, and the corresponding
quantity on the electrolyte side, $q_\mathrm{e}$. (By the
requirement of charge neutrality, $q_\mathrm{e} = - q_\mathrm{met}$.)
Considering $q_\mathrm{met}$ as a function of $\theta$ and $E$, 
one thus gets the current density 
\begin{equation}
i = \frac{\mathrm{d} q_\mathrm{met}}{\mathrm{d} t}
= \left( \frac{\partial q_\mathrm{met}}{\partial \theta} \right)_E 
  \frac{\mathrm{d} \theta }{\mathrm{d} t}
+ 
\left( \frac{\partial q_\mathrm{met}}{\partial E} \right)_\theta 
  \frac{\mathrm{d} E }{\mathrm{d} t}
\;.
\label{eq:i}
\end{equation}
From the electrocapillary equation for the compact double layer 
(i.e., the full differential of the surface free energy density
or ``surface tension" $\sigma$),
\begin{equation}
- \mathrm{d} \sigma = 
(z_\mathrm{met} \Gamma_\mathrm{met} - \Gamma_{e^-} + \lambda \theta) e 
\, \mathrm{d} E + \theta \mathrm{d} \mu_\mathrm{S} 
+ \sum_j \Gamma_j \mathrm{d} \mu_j 
\;,
\label{eq:elcap}
\end{equation}
they obtain 
\begin{equation}
q_\mathrm{met} \equiv  
\left( \frac{\partial \sigma}{\partial E} \right)_{\mu_\mathrm{S},\mu_j} 
=
(z_\mathrm{met} \Gamma_\mathrm{met} - \Gamma_{e^-} 
                  + \lambda \theta) e
\;,
\label{eq:qm}
\end{equation}
where $z_\mathrm{met}$, $\Gamma_\mathrm{met}$, and $\Gamma_{e^-}$ refer
to the ions and conduction electrons of the electrode. 
The current density {\it at constant potential\/},
$i_\mathrm{pot}$, is thus obtained from Eq.~(\ref{eq:i}) by 
setting $\mathrm{d} E / \mathrm{d} t = 0$.
The necessary relation to $\gamma$ as defined in Eq.~(\ref{eq:gam})
is found by writing
\begin{equation}
\left( \frac{\partial q_\mathrm{met}}{\partial \theta} \right)_E 
= \left( \frac{\partial q_\mathrm{met}}{\partial \mu_{\mathrm S}} \right)_E
\left( \frac{\partial \mu_\mathrm{S}}{\partial \theta} \right)_E
\label{eq:chain}
\end{equation}
and using the Maxwell relation obtained from the electrocapillary equation,
\begin{equation}
\left( \frac{\partial q_\mathrm{met}}{\partial \mu_\mathrm{S}} \right)_{E}
=
\left( \frac{\partial \theta}{\partial E} \right)_{\mu_\mathrm{S}}
\;,
\label{eq:max}
\end{equation}
together with the standard equality for any three quantities
related by a single equation, 
\begin{equation}
\left( \frac{\partial \theta}{\partial E} \right)_{\mu_\mathrm{S}}
\left( \frac{\partial E}{\partial \mu_\mathrm{S}} \right)_{\theta}
\left( \frac{\partial \mu_\mathrm{S}}{\partial \theta} \right)_{E}
= -1
\;.
\label{eq:std}
\end{equation}
Thus,
\begin{equation}
i_\mathrm{pot} = 
\left( \frac{\partial q_\mathrm{met}}{\partial \theta} \right)_E
\frac{\mathrm{d} \theta }{\mathrm{d} t}
= - \gamma e \frac{\mathrm{d} \theta }{\mathrm{d} t}
  \;.
\label{eq:ipot}
\end{equation}

To appreciate this derivation it is important to realize that the
current at constant potential is not a simple quantity.
As pointed out by Schmickler \cite{SCHM88}, 
conceptually it involves two steps: first the adsorption of the ion
and corresponding buildup of the image charge, {\it which changes
the potential\/}, followed by readjustment of the charges to bring
the potential back to its original value. 
Without a detailed, microscopic model, 
these semimacroscopic, thermodynamic results cannot tell us
in detail how the charges are distributed in the interface region. 
This is a major limitation of the concept of electrosorption valency. 

The  applicability of the results summarized above 
to reversible electrodes (as opposed to
perfectly polarizable ones) was questioned by Frumkin, Damaskin,
and Petrii in Ref.~\cite{FRUM74}. 
In response, Vetter and Schultze explicitly established  
the validity of their results for reversible electrodes 
in Ref.~\cite{VETT74}.

\section{Surface Dipole Moment}
\label{sec:P}

While the quantities discussed in Sec.~\ref{sec:Q} are purely
thermodynamic quantities (in the case of excess supporting
electrolyte), a microscopic theory of the interface structure is
necessary to estimate the dipole moment associated with adsorption
of an ion \cite{SCHM88,SCHM87,FORE96}. 
Generally, the dipole moment of a (one-dimensional) 
charge distribution $q(x)$ is defined as 
\begin{equation}
p = \int_{x_1}^{x_2} x \, q(x) \, \mathrm{d} x
\;.
\label{eq:dip}
\end{equation}
The result is independent of the coordinate system if 
the integration limits are chosen such that charge neutrality
is obeyed over $[x_1,x_2]$: 
$\int_{x_1}^{x_2} q(x) \, \mathrm{d} x = 0$ \cite{JACK75}. 
The charge distribution produces a potential difference, $\phi_2 -
\phi_1 = p/\epsilon$, where $\epsilon$ is the dielectric constant
of the medium. The details depend on the microscopic model. 

The approximation used in Ref.~\cite{HAMA05B} is the commonly used
one \cite{WAND01,WASI02B}, due to Bange et al.\ \cite{BANG87}
and Schmickler \cite{SCHM96,SCHM88}, 
\begin{equation}
p = \frac{z e \epsilon}{C_\mathrm{H}} 
\left(1 - \frac{\gamma}{z} \right) \;,
\label{eq:dip2}
\end{equation}
where $C_\mathrm{H}$ is the Helmholtz capacity. 
This result is derived under the condition of charge neutrality
(but not necessarily vanishing dipole moment) over the DDL. It can be
obtained simply within the picture illustrated in Fig.~\ref{fig:one}
as follows. The dipole moment is related to the potential drop from
the metal to the layer of specifically adsorbed ions as 
$p = \epsilon (\phi_\mathrm{met} - \phi_\mathrm{IHL}) 
= \epsilon (1-g) E$. Assuming the field dependence of
$\gamma$ in Eq.~(\ref{eq:gam2}) can be included in $g$ and $\lambda$,
and ignoring the polarization terms $\kappa$ when solving  
Eq.~(\ref{eq:gam3}) for $(1-g)$, we obtain 
\begin{equation}
p = \frac{z \epsilon E}{z+\lambda}\left( 1 - \frac{\gamma}{z} \right)
\;.
\label{eq:dip3}
\end{equation}
By setting $C_\mathrm{H} = e (z+\lambda)/E$, we get 
Eq.~(\ref{eq:dip2}), which can be rewritten in terms of $(1-g)$ and
the partial charge-transfer coefficient $\lambda$ as 
\begin{equation}
p = \frac{z e \epsilon}{C_\mathrm{H}} 
\left( 1-g \right) \left(1 + \frac{\lambda}{z} \right) 
\;.
\label{eq:dip4}
\end{equation} 
The prefactor $z e \epsilon / C_\mathrm{H}$
can be viewed as an effective {\it dipole distance\/}. However, due to
screening by both the liquid and the electron gas of the metal,
this distance is generally much smaller than the ionic radius of the
adsorbate \cite{SCHM88,SCHM87}. Equation~(\ref{eq:dip4}) with $g=0$
was the one used in Ref.~\cite{HAMA05B}
for the dipole moment, and therefore for the lateral adsorbate 
interactions. 

An alternative approach is to consider the surface dipole moment as the
basic, physical quantity, and $\gamma$ as a derived quantity linked to
$p$ by Eq.~(\ref{eq:dip2}) under assumption of the geometric model
described in Fig.~\ref{fig:one}.

\section{Conclusion}
\label{sec:C}

In this discussion paper we have summarized arguments showing that 
the electrosorption valency
$\gamma$ as defined by Vetter and Schultze \cite{VETT72A} gives the
correct result for the current in the external circuit due to
adsorption of an ion of charge $ze$, Eq.~(\ref{eq:ipot}) \cite{VETT72B}, 
as well as the
relation between $\gamma$ and the charge-transfer coefficient
$\lambda$, Eq.~(\ref{eq:gam3}) \cite{VETT72A}. These results show that
there is no need for coadsorption with an oppositely charged ion to
reduce the current per adsorbate particle from $ze$, to
$\gamma e$. We note that the derivation of these results requires
charge neutrality over the whole interface region, as indicated in
Fig.~\ref{fig:one}, not separately over the electrolyte and the
compact double layer, as we understand the argument of LH to
imply. We believe the reason for their conclusion that partial
charge transfer cannot change the current is a result of their
overly strict charge-neutrality requirement. 

Beyond these general results, we have shown that the approximations
used in Refs.~\cite{HAMA05B,HAMA03} are excess supporting electrolyte
and vanishing of the factor $g$. While certainly
not exact, these assumptions are reasonable for the systems studied
and give very good
agreement between the computer simulations of the lattice-gas model
and the experimental adsorption isotherms. 

Finally we note that the arguments given here are based on
classical thermodynamics with no specific assumptions on the
microscopic structure of the interface region, beyond the charge
neutrality. In order to obtain explicit results for microscopic
parameters without fitting to experiments, one would need
quantum-statistical-mechanical calculations that are still beyond
our computational capacity. Only such future calculations 
have the potential to determine
explicitly such quantities as the surface dipole moment and the
spatial distribution of charge and dipole moments in the whole interfacial
region.

\section*{Acknowledgments}

We appreciate comments on the manuscript by S.~Frank. 

This work was supported in part by U.S.\ National Science
Foundation Grant No.\ DMR-0240078, by Florida State University
through its Center for Materials Research and Technology and its 
School of Computational Science, and by Research Centre J{\"u}lich.










\clearpage

\end{document}